\documentclass[%
 aip,
 jmp,%
 amsmath,amssymb,
 reprint,%
]{revtex4-2}
\usepackage[version=4]{mhchem} 
\usepackage{color}

\newcommand{\blue}{\textcolor{black}}
\usepackage{graphicx}
\usepackage{dcolumn}
\usepackage{bm}

\begin{document}
\title[Beyond Lambertian light trapping]{Beyond Lambertian light trapping for large-area silicon solar cells\blue{: fabrication methods}
}
\author{Jovan Maksimovic}
\author{Jingwen Hu}\thanks{J.M. and J.H. contributed equally.}
\author{Soon Hock Ng}
\author{Tomas Katkus}
\author{\\ Gediminas Seniutinas}
\affiliation{Optical Sciences Centre and ARC Training Centre in Surface Engineering for Advanced Materials (SEAM), School of Science, Swinburne University of Technology, Hawthorn, Vic 3122, Australia.}
\author{Tatiana Pinedo Rivera}
\author{Michael Stuiber}
\affiliation{Melbourne Centre for Nanofabrication, ANFF Victoria, 151 Wellington Rd., Clayton Vic 3168 Australia}
\author{\\ Yoshiaki Nishijima}
\affiliation{Department of Electrical and Computer Engineering, Graduate School of Engineering, Yokohama National University, 79-5 Tokiwadai, Hodogaya-ku, Yokohama, 240-8501, Japan}
\affiliation{Institute of Advanced Sciences, Yokohama National University, 79-5 Tokiwadai, Hodogaya-ku, Yokohama 240-8501, Japan}
\author{Sajeev John}
\affiliation{Department of Physics, University of Toronto, 60 St. George Street, Toronto, ON, M5S 1A7, Canada}
\author{Saulius Juodkazis}\thanks{For correspondence: S.Jo. and S.Ju.}
\affiliation{Optical Sciences Centre and ARC Training Centre in Surface Engineering for Advanced Materials (SEAM), School of Science, Swinburne University of Technology, Hawthorn, Vic 3122, Australia.}
\affiliation{World Research Hub Initiative (WRHI), School of Materials and Chemical Technology, Tokyo Institute of Technology, 2-12-1, Ookayama, Meguro-ku, Tokyo 152-8550, Japan} 
\email{John@physics.utoronto.ca; SJuodkazis@swin.edu.au}

\date{\today}

\begin{abstract}
Light trapping photonic crystal (PhC) patterns on the surface of Si solar cells provides a novel opportunity to approach the theoretical efficiency limit of 32.3\%, for light-to-electrical power conversion with a single junction cell. This is beyond the efficiency limit implied by the Lambertian limit of ray trapping $\sim$29\%. The interference and slow light effects are harnessed for collecting light even at the long wavelengths near the Si band-gap. We compare two different methods for surface patterning, that can be extended to large area surface patterning: 1) laser direct write and 2) step-\&-repeat $5^\times$ reduction projection lithography. Large area throughput limitations of these methods are compared with the established electron beam lithography (EBL) route, which is conventionally utilised but much slower than the presented methods. Spectral characterisation of the PhC light trapping is compared for samples fabricated by different methods. Reflectance of Si etched via laser patterned mask was $\sim 7\%$ at visible wavelengths and was comparable with Si patterned via EBL made mask. The later pattern showed a stronger absorbance than the Lambertian limit. (M.-L. Hsieh et al., Sci. Rep. 10, 11857 (2020)).      
\end{abstract}

\keywords{Silicon solar cells, laser ablation, light trapping, Lambertian limit}
\maketitle


\section{Introduction}
\label{intro}  
\begin{figure}[t]
\centering\includegraphics[width=1\linewidth]{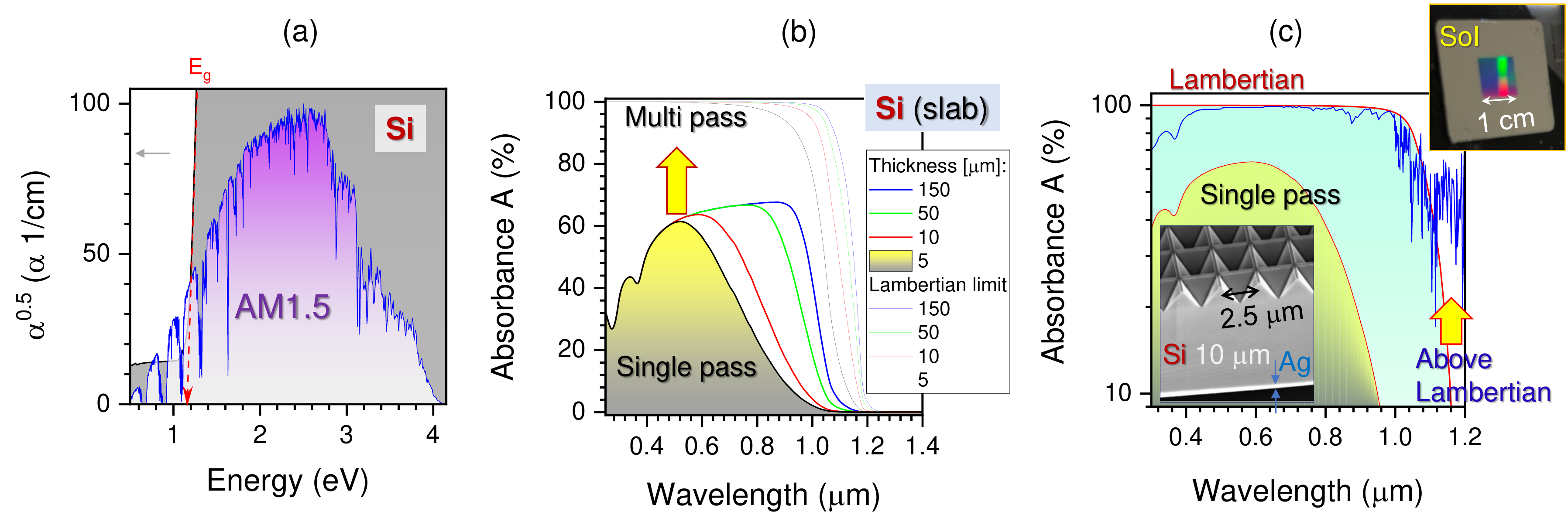}
\caption{Why Si solar cells? (a) Absorption coefficient $\alpha$ of Si vs photon energy (black line, grey shading) and solar spectrum profile (blue line, purple shading) at air mass AM1.5 ground conditions~\cite{am15}. The arrow points to the bandgap energy $E_g$. (b) Calculated absorbance for a Si slab with different thickness in single and multiple (the Lambertian limt) passes. (c) Absorbance of 10-$\mu$m-thick Si (device layer of SOI): single pass, Lambertian limit, and calculated PhC light trapping for the used geometry~\cite{Hsieh}. Inset shows the PhC structure made on a Si-on-insulator (SOI) substrate which exceed the ray optics Lambertian limit~\cite{Hsieh}. 
} \label{f-conc}
\end{figure}

Solar installations will soon provide 1~TW of power from $\sim 16$~TW we consume per year now with an annual uptake of solar increasing by $\sim 35\%$~\cite{Wilson2020,PV}. Around the world, the most common type of photovoltaic installation is the venerable silicon solar cell, which has a typical efficiency of $\sim 18\%$\blue{~\cite{Kopidakis}}.
The efficiency of a hypothetical Si solar cell which perfectly absorbs all sunlight in the 300-1200~nm wavelength range and has no bulk or surface recombination losses is defined by the Shockley–Queisser limit of $32.3\%$ at room temperature~\cite{Shockley1961}. Real solar cells cannot absorb all such sunlight and even the purest silicon suffers from Auger recombination losses. The ray optics picture of light trapping suggests that solar absorption cannot exceed the so-called Lambertian limit. To go beyond the ray optics limit, a wave-interference-based light trapping solution needs to be engineered~\cite{SJ19} (\textbf{Figure~\ref{f-conc}}(a) and \ref{f-conc}(b)). The Lambertian limit is given by the maximum of optical path enhancement in a weakly absorbing and optically thick medium $F = 4n^2$, here $n$ is the real part of the complex refractive index $\Tilde{n} = n +i\kappa$. The Lambertian limit of absorbance $A_L = \alpha d/(\alpha d + 1/F)$, with $\alpha = 4\pi\kappa/\lambda$ being the absorption coefficient at the wavelength $\lambda$ and $d$ is the thickness of the cell.  

It was recently shown that photonic crystals (PhC) made via surface texturing and targeting the 0.9-1.2~$\mu$m spectral range is the key to surpassing the light trapping Lambertian limit~\cite{Hsieh}. 
A thin 10-$\mu$m device layer of silicon-on-insulator (SOI) was textured 
by inverted pyramid or tee-pee PhC structures of lattice constant $\Lambda = 2.5~\mu$m or 1.2~$\mu$m, respectively (Figure~\ref{f-conc}(c)). The demonstrated light trapping results propels the solar-to-electrical energy conversion to 
$\sim 31\%$~\cite{Hsieh}. This 
performance was predicted theoretically and 
it is attainable with only $\sim 10$-$15~\mu$m-thick Si~\cite{30}, which is very attractive for applications \blue{of lightweight flexible solar cells that can be coated on a variety of surfaces and have higher power conversion efficiency than present-day solar panels~\cite{simovski_tretyakov_2020}. Flexible multi-junction solar cells are sought after in the automotive industry where high efficiency is required and there is a limitation of available surface area~\cite{Satou}.} 

Interdigitated back contacts (IBC) solar cells 
are currently the most efficient, breaking through the long-held 25\% efficiency and performing above $26\%$, closing the gap to the Lambertian limit~\cite{Hollemann, Green2009}. 
The final 
selection of an optimised 
surface texturing 
will be made upon full characterisation of the material's response~\cite{Hollemann}. Recently, it was shown 
that surface recombination velocity (SRV) in 
IBC Si solar cells can reach 
10~cm/s after processing~\cite{Hollemann}. Surface nano-texturing by 
nanoscale needles - black-Si - can be utilised for anti-reflective function~\cite{16aplp076104}. However, when black-Si was applied to Si solar cells, the 
best efficiency of only $\sim 22\%$ was achieved~\cite{20MTE100539}. Also, a prohibitively high 
SRV value of $\sim 100$~cm/s was estimated on etched surfaces of black-Si~\cite{16semsc221}. It is not expected that the light-trapping methods based on ray-optics, 
when applied to conventional Si solar cell architecture, 
allows it to reach efficiencies beyond 28\%~\cite{SJ19}. 
The optimised IBC solar cells of 15~$\mu$m thickness with PhC light trapping should 
perform at $\sim 30$\% efficiency even when 
SRV is as high as 100~cm/s~\cite{SJ19}. \blue{The importance of good surface passivisation for Si solar cells~\cite{Peibst1} was demonstrated recently by record performing (26\% efficient) both-side-contacted~\cite{Glunz} and SiC-based highly transparent passivating contacts on 24\% efficent cells~\cite{Pomaska}.}

In this study, we compare two different methods to define and fabricate PhC light trapping on Si using etch mask patterning by: 1) direct laser writing by ablation and 2) stepper photo-lithography. For comparison, electron beam lithography (EBL) (as in ref.~\cite{Hsieh}) was also used. Process development and challenges are compared towards the goals of large area 
$\sim 2\times 2$~cm$^2$ fabrication of light trapping textures on 
Si, from the perspective of area up-scaling, throughput and technological bottlenecks.  
\begin{figure}[t]
\centering\includegraphics[width=1\linewidth]{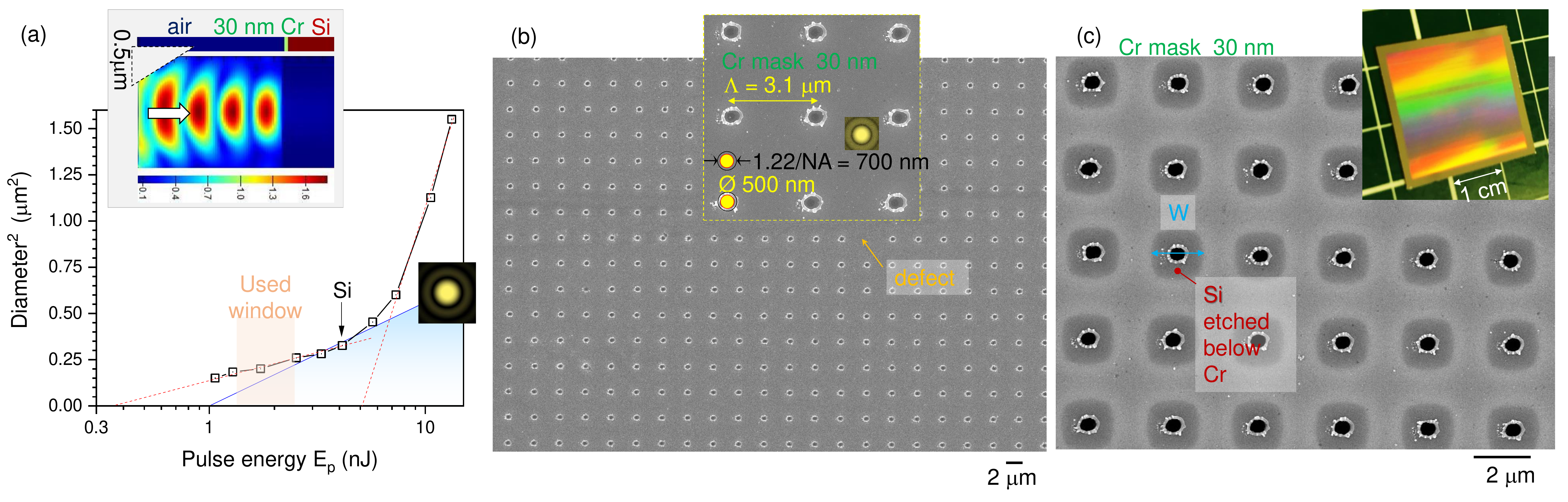}
\caption{Cr etch mask defined by direct laser writing. (a) Calibration of the ablated hole diameter $D^2\propto \ln(E_p)$ vs the pulse energy $E_p$. The threshold 0.36~nJ (0.09~J/cm$^2$, 0.41~TW/cm$^2$) and 5.15~nJ (1.35~J/cm$^2$, 5.85~TW/cm$^2$) are the intersections of dashed-lines with $x$-axis. The solid-line is expected evolution of $D^2$ defined by the beam waist $r = 0.61\lambda/NA$. Top-inset shows light field $E$ simulations for the $NA = 0.9$ focusing used with a Cr film on Si, for $\lambda = 515$~nm; the finite difference time domain (FDTD, Lumerical) was used. (b) SEM image of the ablated 30~nm Cr film; $E_p = 1.7$~nJ. The inset shows a closeup view of the ablated holes in the Cr film; the thumbnail image of the Airy pattern of the focal spot is scale matched. (c) SEM image of sample after $ 10$~min plasma etch. The contrast change reveals the width $W$ of the plasma etched Si (\ce{SF6}:\ce{CHF3}:\ce{O2} at 5:1:1 flow rate ratio). The inset shows photo of $2\times 2$~cm$^2$ area of ablated Cr mask on Si (before etch). } \label{f-Cr}
\end{figure}
\begin{figure}[t]
\centering\includegraphics[width=0.8\linewidth]{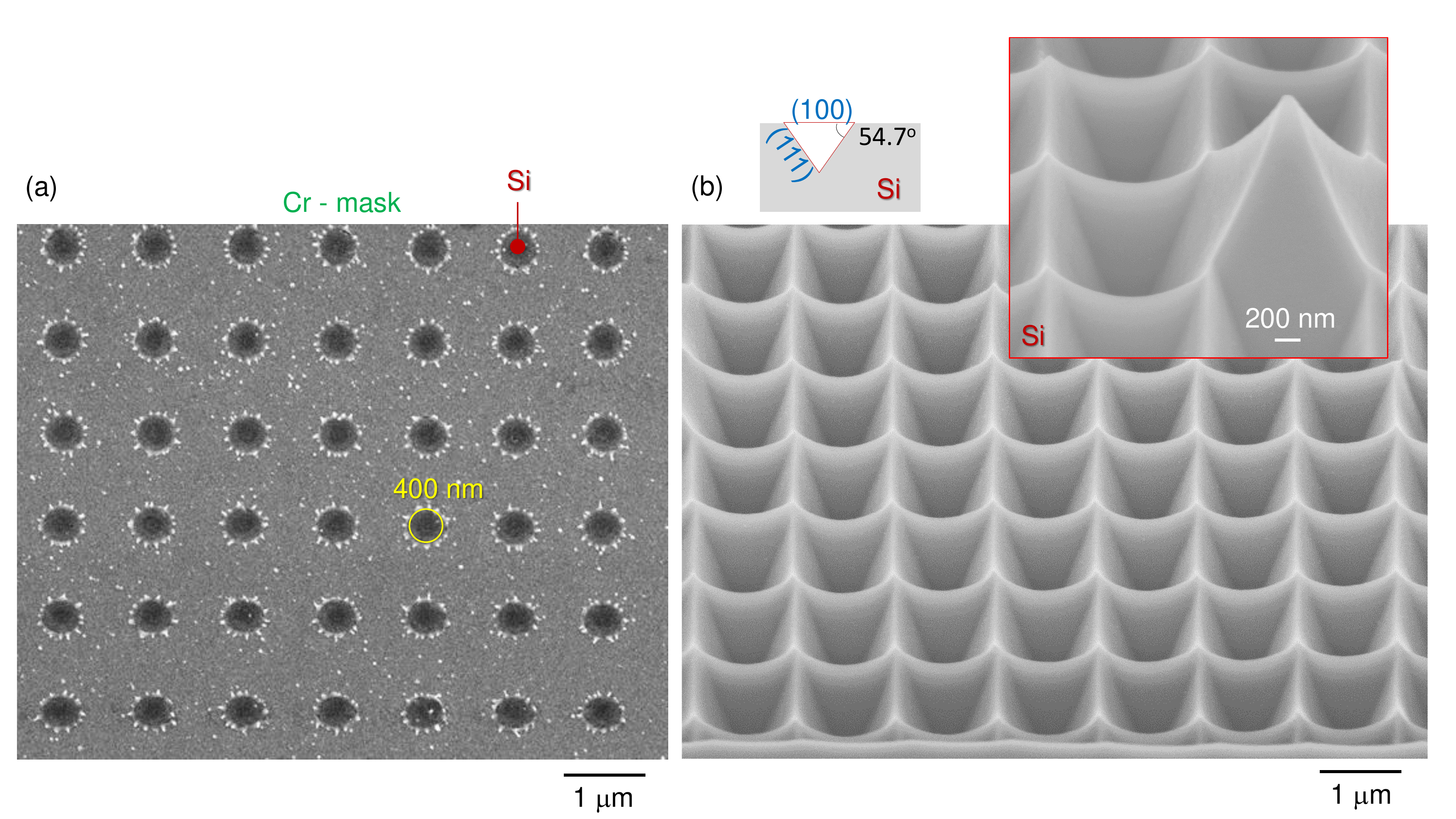}
\caption{SEM images of a hard mask for RIE made in 50-nm-thick Cr film by laser ablation with 515~nm wavelength, 230~fs single laser pulses. Etched pattern after Cr mask removal. The inset shows an under-etch in the region where one ablation opening in the mask was absent. Conditions: $NA = 0.9$ lens was used, write speed 10~cm/s, pulse energy $5$~nJ, laser repetition rate 200~kHz. } \label{f-abla}
\end{figure}
\begin{figure}[b!]
\centering\includegraphics[width=0.85\linewidth]{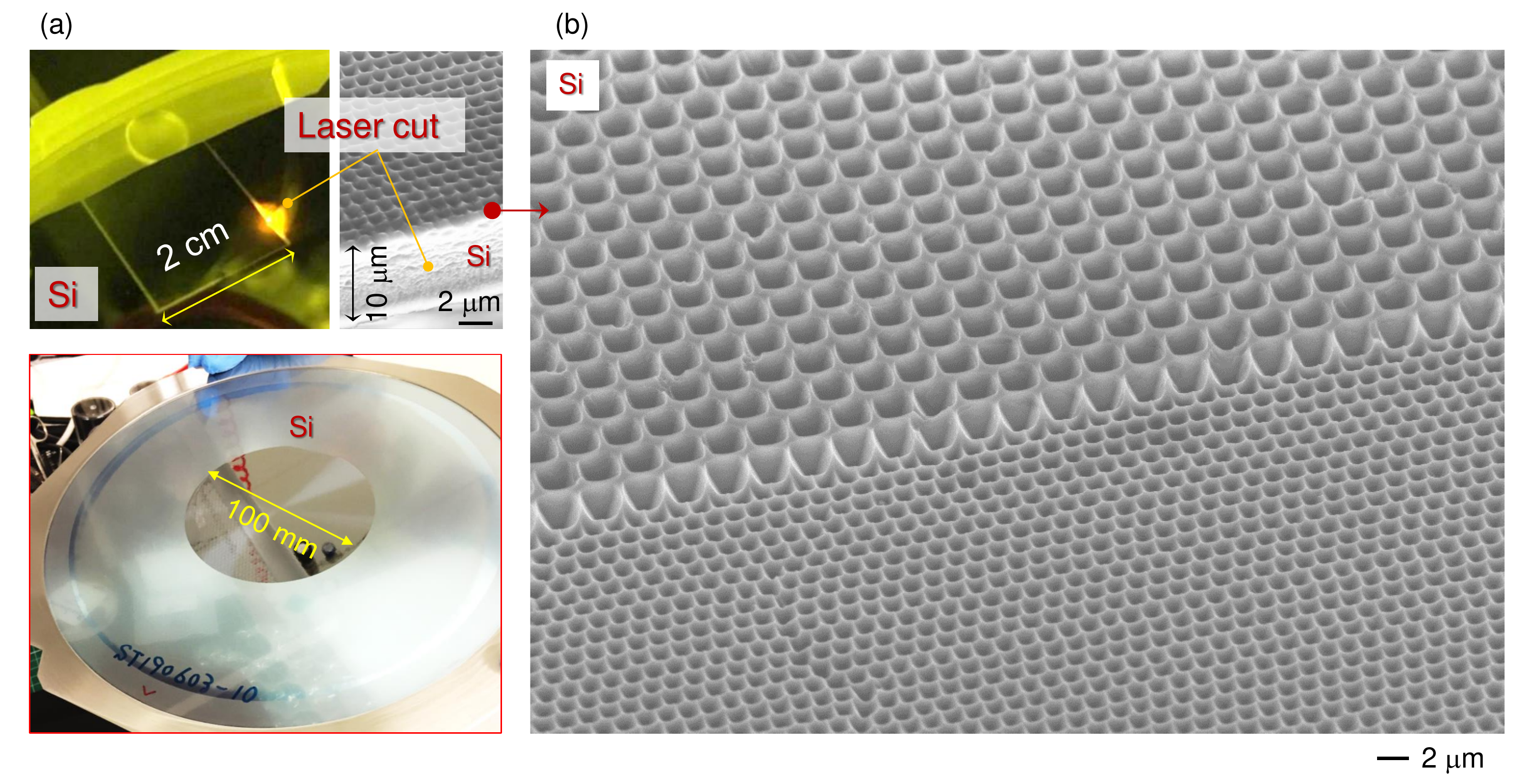}
\caption{(a) 10-$\mu$m-thick Si wafer cut by $E_p\approx 0.8~\mu$J energy fs-laser pulses (1030~nm/230~fs/200~kHz)  using $NA = 0.14$ objective lens at 1~cm/s scan speed (5 passes with focus on the surface). SEM image of the 10-$\mu$m-thick Si wafer after cutting and patterning. Optical image of the 100-mm-diameter Si wafer on a plastic film. (b) PhC surface etched through holes ablated by fs-laser to define the Cr etch mask; $NA = 0.9$ lens was used, write speed 10~cm/s, pulse energy 12~nJ, laser repetition rate 200~kHz. Note the step height between regions with different period $\Lambda$. This step change is due to the under etch which is also responsible for formation of the upward-pyramid shown in inset of Figure~\ref{f-abla}(b). } \label{f-laser}
\end{figure}
\begin{figure}[t]
\centering\includegraphics[width=1\linewidth,angle=0]{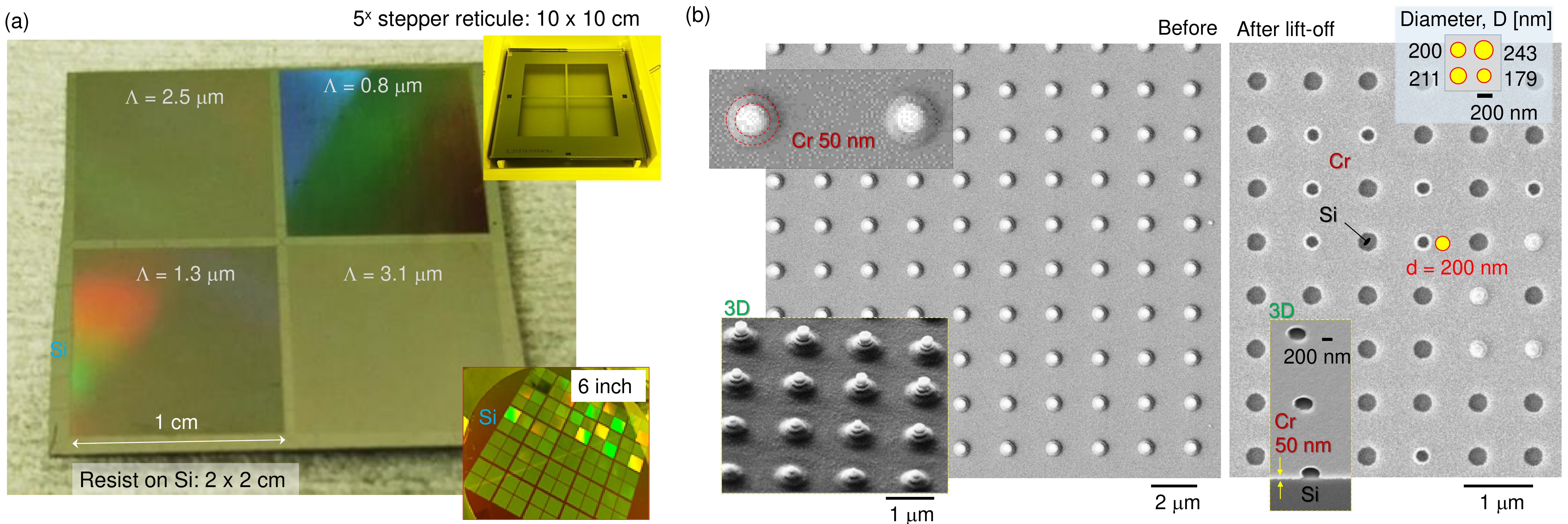}
\caption{(a) Photo of the dot patterns 
(resist on Si) 
after exposure, development and dicing. Insets: (bottom)  
image of the entire 6-inch Si wafer after exposure and development; (top) the stepper reticle used for the i-line (365~nm Hg) $5^\times$ 
reduction 
(positive resist: the exposed regions are removed). (b) SEM images of 
50-nm Cr film e-beam evaporated on
700-nm-thick resist patterns before and after lift-off; exposure was made by projection lithography on Si. Circular marker shows $d = 200$~nm for comparison with developed holes. Bottom-insets shows slanted-view images of the patterns after lift-off. Top-right inset shows the hole openings $D$ for the four segments of PhC patterns with different periods shown in (a).} \label{f-tsukuba}
\end{figure} 

\section{Results}\label{res}

\textbf{Laser ablation of etch mask}. The simplest method to define a mask for plasma-based reactive ion etching (RIE) and/or wet etch of Si is direct laser ablation of sub-1~$\mu$m diameter holes in a film (\textbf{Figure~\ref{f-Cr}}). Chromium, which is known for good adhesion to many different surfaces and is compatible with plasma and wet etch protocols of Si, was e-beam evaporated on Si and used for the mask. The refractive index of Si and Cr at $\lambda = 515$~nm, the wavelength used for ablation, is $\Tilde{n}_{Si} = 4.223 +i0.061$ and $\Tilde{n}_{Cr} = 2.894 +i3.323$, respectively. The reflectance of Si and Cr (see 
Section~\ref{meth}) are $R_{Si} = 38.1\%$ and $R_{Cr} = 55.8\%$, respectively (note, for bulk samples). A 
film of chromium $d = 30$~nm thickness transmits only $T_{Cr} = \exp{(-\alpha_{Cr}d)} = 8.8\%$ of pulse, where 
$\alpha_{Cr} = 4\pi\kappa_{Cr}/\lambda$. The absorbed component 
of the energy $A = 1-R-T$ is utilised 
for ablation. 
This shows that energy is 
deposited almost exclusively into the Cr film, even if it is nano-thin (inset in Figure~\ref{f-Cr}(a)). For focusing with a numerical aperture $NA = 0.9$ objective lens, the focal spot radius was $w_0=0.61\lambda/NA = 349$~nm, the average fluence per pulse $F_p=E_p/(\pi w_0^2) = 0.26$~J/cm$^2$ for $E_p = 1$~nJ and average pulse irradiance (intensity) $I_p=F_p/t_p = 1.14$~TW/cm$^2$ for $t_p = 230$~fs pulse duration. This coincides with 
the ablation threshold of Si $\sim 0.2$~J/cm$^2$~\cite{Mathias}. 

The diameter of the ablated hole $D$ is defined by the pulse energy density (fluence), which for the Gaussian beam $F(r) = F_0\exp(-2(r/w_0)^2)$ follows the $D^2 = 2w_0^2\ln(E_p)$ dependence plotted in Figure~\ref{f-Cr}(a), here $r$ is the radial coordinate of Gaussian fluence (intensity) distribution, $w_0$ is the waist of the beam and $F_0 = 2F_p$ is the amplitude of fluence~\cite{Brendon}. The linearised equation for the $NA = 0.9$ used focusing is plotted by a solid line; see the thumbnail-inset of the Airy pattern where the first minimum is used as an approximation for the waist of the beam. For a bulk sample, this dependence would determine the threshold of ablation at the x-axis intercept. For the nano-film used in this study, two distinct regions with different slopes and thresholds were observed. Changes of the slope occurred when an ablation of Si become evident in the ablation crater (see the arrow marker Si in Figure~\ref{f-Cr}(a)). Different slopes reflect different formation mechanisms of the ablation hole. The smaller threshold of $\sim 0.1$~J/cm$^2$ is consistent with Cr ablation since the most of pulse energy is deposited in Cr as calculated above. The larger threshold for the faster growing part of the dependence reflects thermal diffusion influence in the formation of larger ablation craters. Detailed mechanisms of the formation of the ablated opening in the mask is beyond the scope of the current study. The pulse energy $E_p = 2\pm 0.5$~nJ (on the sample) was chosen for reproducible ablation of $D\approx 0.45\pm 0.10~\mu$m holes; pulse-to-pulse stability of laser was $\sim 1\%$. They were slightly smaller than the diameter of the focal spot as defined by the first minimum of the Airy pattern; see inset in Figure~\ref{f-Cr}(b). 
The ring around the rim of ablated crater facilitated long plasma etch times with no mechanical support from under etched Si. The ring is likely made out of a nano-alloy of Si and Cr which was not etched during the standard Cr-etch (see Supplementary \textbf{Figure~\ref{f-alloy}} for discussion). With the established mask fabrication conditions, it was possible to fabricate large areas ($2\times 2$~cm$^2$ inset in Figure~\ref{f-Cr}(c)) within several hours (up to 10~h dependent on pattern density). The 20-30~nm Cr film is partly transparent in SEM observation and was used to determine the required etch time to obtain the narrowest ridges $R^{'}$ between the etched tee-pees (inverted pyramids); here the period $\Lambda = W + R^{'}$. After plasma or wet KOH etch of PhC pattern on Si, Cr mask was removed with a chrome etch solution.    

A detailed summary of the width $W$  evolution during etch time for different diameter holes in Cr-mask ablated by correspondingly different single laser pulses of energy $E_p$ are presented in \textbf{Figure~\ref{f-plasma}}. We took advantage of the semi-transparent nature of the 30~nm Cr mask in the SEM to explore structure changes during etching. For pulse energies $\sim 1$~nJ, Si:Cr alloying took place instead of ablation (see Supplement for details). This was evidenced during etching when those regions acted as a mask rather than a hole. When pulse energies exceeded $\sim 8$~nJ, less regular ablation craters were observed, which showed crack development in Cr mask after long $> 10$~min etching times. Under etch of Si via ablated hole in the Cr mask was slow and followed slope $W\propto E_p^\gamma$ with $\gamma <1/2$ (Figure~\ref{f-plasma}). Silicon under etch below the Cr-mask showed an anisotropy guided by the crystallinity of the $(100)$ Si plane, which is most pronounced in the case of wet etching in KOH when inverted pyramids are formed. This anisotropy was responsible for the formation of under etch which was slowest in the diagonal direction of tee-pees and caused formation of protruding corners (marked as cones in Figure~\ref{f-plasma}(b)). Interestingly, the metal masks are known to facilitate under etch even for amorphous substrates~\cite{aniso} and could be used for more intricate control of the pattern. Once neighbouring tee-pee pits separated by $\Lambda = 3~\mu$m approached each other, the narrow ridges were formed (Figure~\ref{f-plasma}(b)). For longer etching, PhC patterns became deeper, however, the depth $H$ and width $W$ of individual tee-pees did not change. This condition was further exploited for the narrowest ridges $R^{'}$. Tens of minutes was enough to etch PhC structures. We used a slightly thicker Cr film of 50~nm for longer plasma etch to ensure complete under etch and the narrowest ridges in earlier studies of phenomena presented (\textbf{Figure~\ref{f-abla}}). The under etch is illustrated by a pyramidal tip which formed in the place of missing ablation hole in the mask (as shown in Figure~\ref{f-Cr}(b)). Patterns of tee-pee PhC etched in the case of under etch showed similarity to those obtained by anisotropic wet etchning in KOH. Wet etching through ablated holes of $D\approx 0.5~\mu$m was taking $\sim 14$~hours in 30\% KOH at $45^\circ$C. Plasma etching took only tens of minutes for similar PhC patterns.

\blue{A dielectric alumina \ce{Al2O3} film of 20-30~nm, which serves as a part of an antireflection and passivation coating of IBC cells~\cite{Hollemann}, can be used for fabrication of PhC patterns on Si using the same laser ablation method demonstrated for Cr at slightly larger laser pulse energy (Fig.~\ref{f-alumina}). A simpler solvent-based ultrasonic bath removal of the alumina mask after plasma etching is another advantage of this processing, which avoids metal-ion wet chemistry.}

Finally, to show capability to process thin Si wafers of $10~\mu$m thickness which is the target for a PhC enhanced light trapping, the same laser processing of Cr mask and plasma etching was carried out \textbf{(Figure~\ref{f-laser})}. Si $2\times 2$~cm$^2$ samples were laser cut from the micro-thin wafer using the same fs-laser with focusing objective with $NA = 0.14$ (Figure~\ref{f-laser}(a)). For handling, the thin Si film was attached to a 4-inch sapphire wafer using resist, AZ1518. The resist is spin coated to be $1.5~\mu$m on the sapphire wafer which acts as an initial carrier substrate, with the resist acting as an intermediary layer between the thin Si and the sapphire. This can be subsequently handled by laser cutting the Si to a desired size (Figure~\ref{f-laser}(a)) and later put into a solution with developer for transfer to a new carrier substrate in the same way as previously mentioned for PhC patterning. It is important to have a very uniform resist film, to keep the attached Si as flat as possible making fabrication conditions ideal. Strong under etch below the Cr mask defined smaller ridges $R^{'}$ for smaller period $\Lambda$ for the PhC as compared with larger $\Lambda$ (Figure~\ref{f-laser}\blue{(b)}). The under etch is manifested by different heights of the PhC patterns with different periods $\Lambda = 3~\mu$m and 1.5~$\mu$m which were laser ablated on Cr mask side-by-side.   

\begin{figure}[t!]
\centering\includegraphics[width=0.9\linewidth]{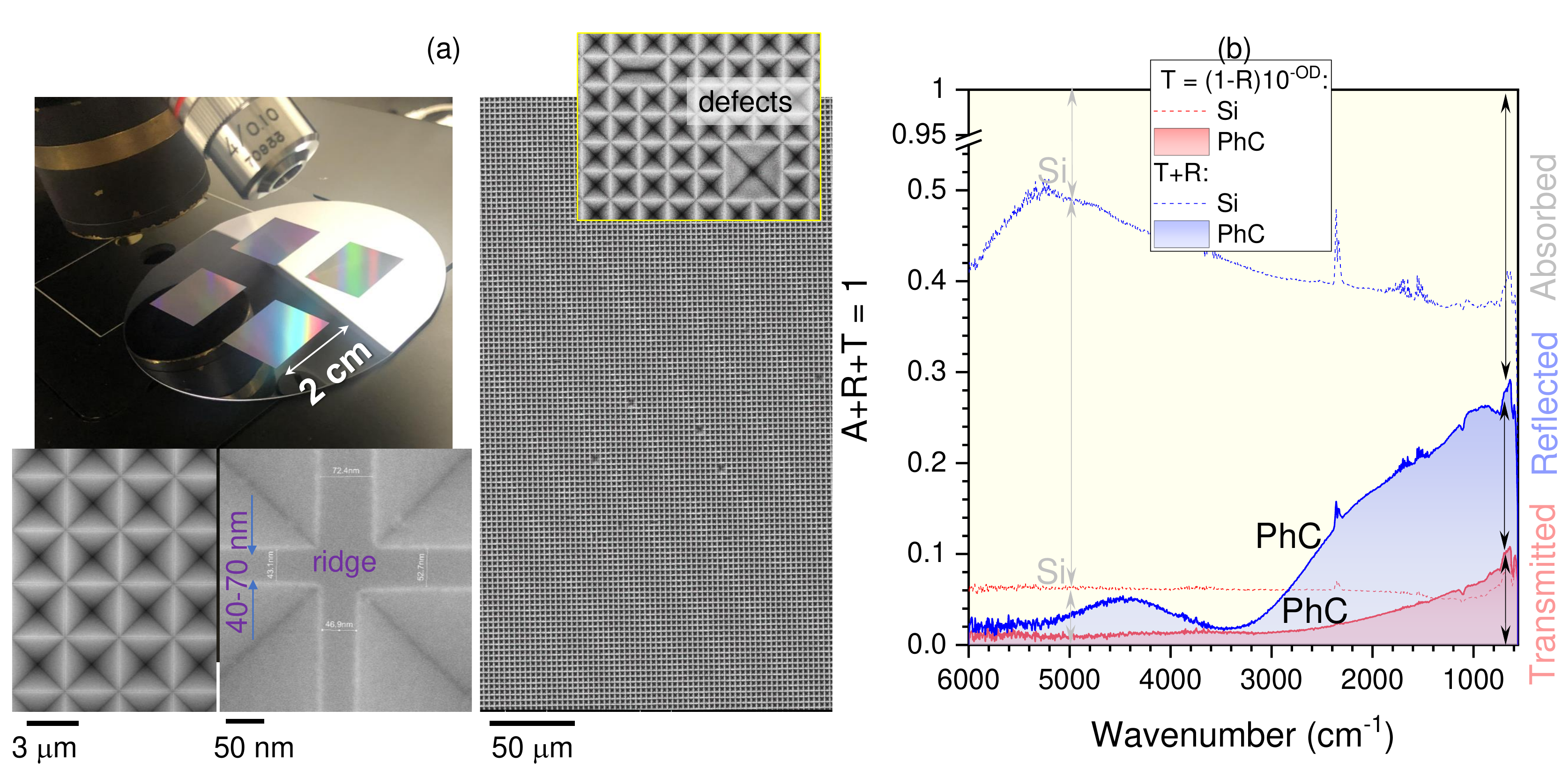}
\caption{(a) Photo of large 2$\times$2~cm$^2$ light trapping PhC patterns 
KOH etched through a chromium-mask defined by EBL 
; period $\Lambda = 3.1~\mu$m. SEM images of the final PhC pattern on 500-$\mu$m-thick n-type Si(100) wafer. (b) Color-coded presentation of portions of the absorbed $A$, reflected, $R$ (measured) and transmitted $T=(1-R)10^{-OD}$ (measured) light at different wavelengths from 1.67 to 25~$\mu$m (or 6000-400~cm$^{-1}$ in wavenumbers $\Tilde{\nu} = 1/\lambda$).} \label{f-2021}
\end{figure}

\textbf{Stepper patterning of etch mask}. 
Steppers reduce and project patterns from a reticle onto a silicon wafer. A $5^\times$ reduction means that the reticle does not require high-resolution electron beam lithography for it's fabrication, even when the resultant pattern is sub-micrometer.
The smallest size 
on the reticle was a 2-$\mu$m-diameter disc, which will produce a 400~nm pillar (ideal conditions) on the etch mask for the Si solar cell. We used Sumiresist PFI-38 photoresist spin coated over 
30-50~nm thickness Cr coatings (e-beam evaporated) for mask fabrication. After fabrication (a wet etch) the 
diameters of the holes in Cr-mask ranged from 180~nm to 245~nm with uncertainty $\sim 10$~nm (\textbf{Figure~\ref{f-tsukuba}}). Patterns with 
smaller period $\Lambda$ experienced a smaller UV exposure dose 
through the mask, which resulted in 
less proximity exposure (Figure~\ref{f-tsukuba}(b)). Top inset shows the actual shape of chromium-coated polymer pillars. 
Consequently, less exposure through optically blocked areas in reticle 
occurred and resulted in the larger diameter pillars 
obtained after development. This difference is 
illustrated by measuring the diameter of the opening in the chromium-mask 
after the lift-off (hole is formed in the location of the resist pillar). The hole diameters and their dependence on the period of the PhC pattern are shown in the inset of Figure~\ref{f-tsukuba}(b). To achieve $\sim$200~nm holes in a 50~nm chromium film is evidence of the high resolution fabrication achieved for patterns defined by stepper photo-lithography. The critical fabrication 
step was lift-off. 
Pillars taller than $500$~nm were required for successful lift-off of the 30-50~nm chromium film (see insets in Figure~\ref{f-tsukuba}(b)). 
Lift-off was not successful for smaller pillars, resulting in missing nano-holes in the mask. 

\textbf{Electron beam definition of etch mask}. We used conventional Si-wafers 
since IBC cell geometry is not compatible with SOI layer stacks. Cr masks were patterned by EBL to define PhC pattern with subsequent wet KOH etching. \textbf{Figure~\ref{f-2021}}(a) shows final PhC with $\Lambda = 3.2~\mu$m made on four regions. Chromium coating 
helped with charge removal and surface positioning (height) for EBL writing. Small ridges $R^{'}$ = 40-80~nm between neighboring inverted pyramids were obtained for longer KOH etch, however, more defects were observed when 
adjacent
pyramids 
merged due to localised under etch. 
The number of defects was between 0.1 to 0.9\%, calculated per $10^3$ pyramids. Figure~\ref{f-2021}(b) shows optical near-IR to IR performance of a $500~\mu$m thick Si wafer patterned with the light trapping PhCs. 
The portions of reflected and transmitted parts of light intensity are from direct measurement and the absorbed portion is calculated from energy conservation 
$A = 1 - R - T$. The PhC surface pattern caused a spectrally wide anti-reflection effect which contributed to the larger portion of absorbed energy. This evident from the difference between a patterned vs blank wafer spectra. Such surface can find applications in optical sensors (strong absorption) and emitters in the IR spectral range~\cite{19sr8284}. 

\begin{figure}[t]
\centering\includegraphics[width=1\linewidth]{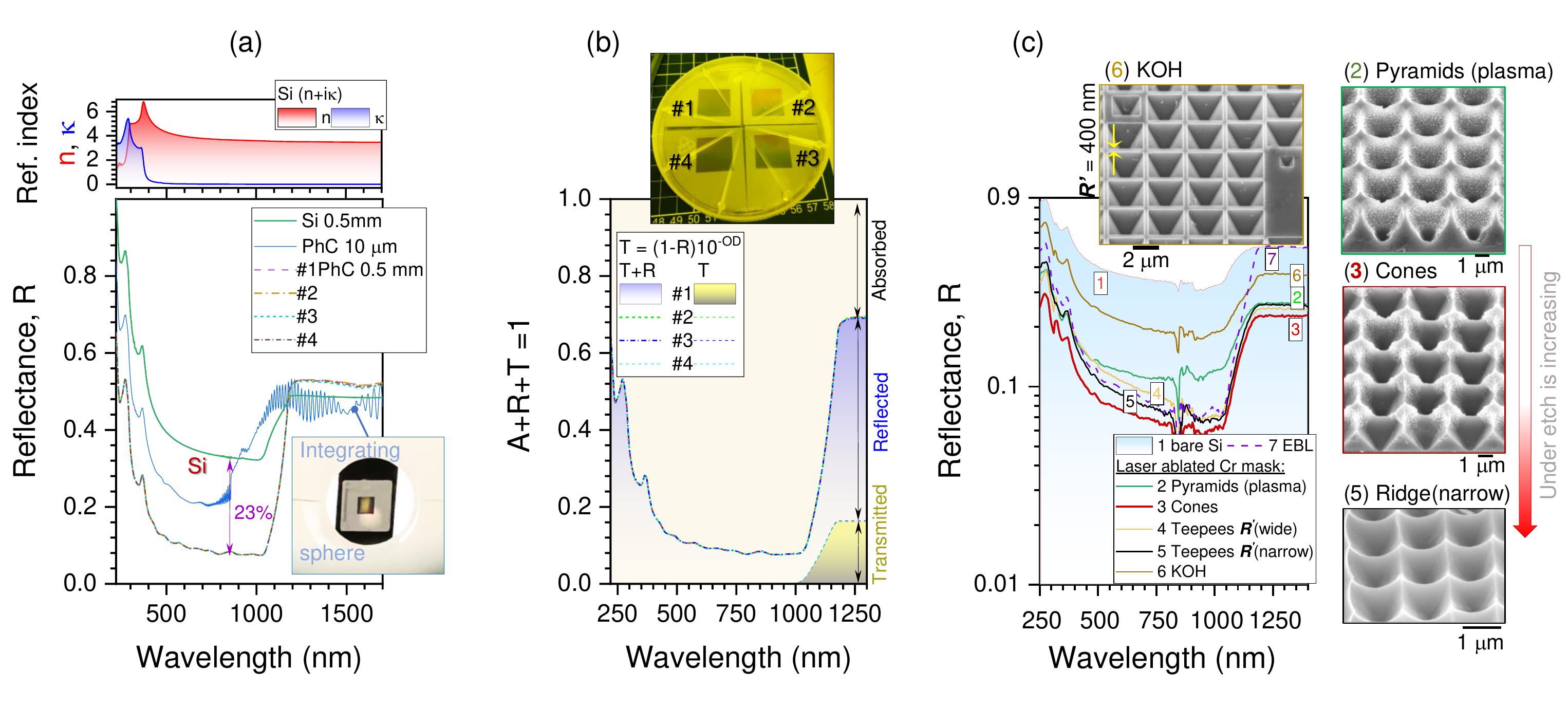}
\caption{Spectral characterisation of PhC light trapping. (a) Reflectance $R$ measured with an integrating sphere for a Si wafer, PhC on SoI (see Figure~\ref{f-conc}(c)), and PhC defined by EBL. Note all four segments (top inset in (b)) showed the same (overlapping) spectral profile. Top inset shows $(n+i\kappa)$ components~\cite{Green2008}. Interference fringes of SoI sample are defined by Si thickness $t = \lambda_1\lambda_2/[2n(\lambda_2 - \lambda_1)]\approx 11~\mu$m (near-IR; $n = 3.46$).  (b) Portions of reflected $R$ (measured), transmitted $T$ (measured) and absorbed $A = 1-R-T$ portions. Inset shows a photo of the 4-inch wafer with 4 EBL defined PhCs. (c) Reflectance of different PhC patterns prepared by laser writing and EBL which were either plasma and wet etched (note $\lg$-axis for $R$). The light spot size on the sample was $\sim 2-3$~mm in diameter in order to reproduce the same conditions on samples with different areas (see the inset in (a) showing sample illuminated inside the measurement chamber). 
} \label{f-spectra}
\end{figure}

\textbf{Figure~\ref{f-spectra}} shows a summary of the spectral characterisation made of different samples in this study; reflectance from the stepper lithography prepared samples was reported in ref.~\cite{Maksimovic2021}, where the highest aspect ratio patterns $H/W\approx 1$ showed the lowest reflectivity. Similar tendency was observed for the PhCs defined by the laser ablated and EBL defined masks (Figure~\ref{f-spectra}). Figure~\ref{f-spectra}(a-b) panels show high repeatability of $R$ spectra measured from four different regions of PhCs defined via EBL (a perfectly overlapping spectra). Reflectivity below $\approx$8\% was obtainable for PhC defined by different methods as long as ridges were narrow and width of the tee-pee $W\rightarrow\Lambda$. The least reflective surface was observed from PhC structures which had corner regions between neighbouring tee-pee pits reaching the level of original Si surface (below Cr mask; see Figure~\ref{f-plasma}(b)). They were among the deepest structures with $h\approx 2.5~\mu$m (marked as cones in Figure~\ref{f-spectra}(c)). The tendency of reduction of reflectance $R$ with under etch, which also caused narrowest ridges $R^{'}\rightarrow 0$, is apparent (see, SEM thumbnail images in (c)). When inverted pyramids were etched by wet KOH or dry plasma processing (\emph{6, 2} in (c)), $R>10\%$ was observed due significant contribution of original Si surface (at ridges) to reflectivity. Application of an antireflective coating for the final solar cell will provide further reduction of reflectance $R<1\%$ and is required passivisation for the reduction of the surface recombination rate. The most promising samples for the practical solar cells will be those with strongly under etched tee-pees due to their low reflectivity and maximum of the removed material which was at a Si-mask interface. The used plasma etch of Si was carried out at low-bias and low-power which makes only shallow $\sim 10$~nm surface modification, which is fully recoverable to almost pristine surface quality by different post-processing steps of annealing, \ce{H2}, \ce{O2} treatments, or their combination, dependent on the used plasma chemistry~\cite{Misra,Mu}.  

With optical the $R(\lambda)$ and $T(\lambda)$ measured experimentally, the absorbed part $A(\lambda)=1-R(\lambda)-T(\lambda)$ can be accessed. Then, the maximum achievable photo current density (MAPD) of the solar cell under AM1.5 (Figure~\ref{f-conc}(a)) illumination of intensity $I_{AM1.5}(\lambda)$ can be calculated~\cite{Hsieh}:
\begin{equation}
J_{MAPD}=\frac{e}{hc}\int^{\lambda=1200nm}_{\lambda=300nm}\lambda\times I_{AM1.5}(\lambda)\times A(\lambda)d\lambda,   
\end{equation}
\noindent where the constants are: $e$ is the electron charge, $h$ is the Plank constant, $c$ is the speed of light. It is assumed that each absorbed photon creates a single electron-hole pair. Consequently, the short-circuit current $J_{SC} =  J_{MAPD}$ for an ideal solar cell without losses due to 
surface and bulk recombination.
How to achieve the latest conditions is an another wide field of research, which is already well matured and achieved world record efficiency in Si solar cell performance. Addition of PhC light trapping is a natural next step to improve light-to-electrical power conversion.  

\section{Discussion}

Laser patterning of a Cr \blue{or \ce{Al2O3}} mask has the fundamental advantage of not using vacuum chambers \blue{for the lithography step for pattern definition} and can be up-scaled for large area fabrication. By combining 
high stage scanning speeds (10~cm/s and faster) 
with high repetition rates (200-600~kHz), it was possible to efficiently pattern $1\times 1$~cm$^2$ surface areas within a reasonable $\sim 2$~hours time. 
Beam travel speeds of meters per second are in reach using galvano and polygon scanners.  
Synchronising the stage movement 
with laser repetition rate results in good alignment of the ablated sites 
between neighbouring lines, at expense of a slightly longer writing time. Different Cr thicknesses necessitate 
different pulse energies to open holes in the mask, but 20-60~nm Cr film can be ablated with good control of hole diameter with sub-20~nJ single pulses. Such pulse energies are now available from fs-laser oscillators and regenerative amplifiers are not required.  
Due to the almost 2$^\times$ sized holes ablated by laser writing as compared with stepper photo-lithography defined mask, approximately 2$^\times$ faster plasma etching was observed. Indeed, 4~mins was all that was needed to achieve optimal ridge conditions $R^{'}\rightarrow 0$~nm with 
the same optimised etching recipe for mask made by laser ablation.  

In areas where there were missing 
holes at the laser ablation sites (missed pulse, slight change in spatial laser mode, etc.), an interesting pattern developed after under etch 
(Figure~\ref{f-abla}(b)). During the initial etch, the expected inverted pyramid pattern formed under the mask.
When etching of the PhC pattern continued through to under etch, 
an upward pyramid was formed 
at the location where intended hole was not ablated. 
These unexpected defects are therefore favourable for the creation of deeper PhC patterns since the crystal orientation of the Si precludes aspect ratios much greater than 1. The results suggest that by creating a checkerboard hole pattern instead of a regular array, each inverted pyramid will be surrounded by four regular pyramids. The resultant PhC will have double the spatial frequency.
This type of defect-mode can be intentionally designed to fabricate deeper PhC patterns and will be explored further. 



Future studies will focus on changing the ratio of \ce{SF6} to \ce{CHF3} to affect the anisotropy (difference in etch rate between in x-y and z) during the dry etching~\cite{etch}.
Tilted angle etching of Si was tested~\cite{20oe16012}, 
however, there was a strong difference in depth of the etched pattern at different distances from the electrode. 
Hence, using this method to control the shape of etched pits is less promising for PhC textured surfaces on solar cells even if predicts a close to Lambertian light trapping~\cite{Han}.

With RIE, varying the bias power greatly influences the etch anisotropy, which is important since the under-etch and aspect ratio $H/W$ can be controlled. Anisotropic etching at a larger bias 
creates deeper patterns,  
while isotoropic sideways etching is enhanced by a larger inductively coupled plasma (ICP) power. 
It was found that the bias voltage had little effect on the aspect ratio of inverted pyramids. 
The $\frac{H}{W}\approx 1$ was observed (0.71 expected from Si lattice) and 
anisotropy of the 
Si etching  was the defining property regardless bias used. 

\section{Conclusions and Outlook}

Two different fabrication methods 
for large-area ($2\times 2$~cm$^2$) PhC patterns on Si for light trapping are demonstrated:
1) direct laser write and 2) stepper photo-lithography. They can scale up for light trapping solutions on order of magnitude larger surfaces and 
were compared with EBL, which recently showed close to the Lambertian limit performance over the visible spectral range and exceed it over 
950-1200~nm wavelengths 
when an antireflection coating was applied on SOI samples~\cite{Hsieh}. In this study, large $2\times 2$~cm$^2$ areas are demonstrated using a Cr etch mask on a Si wafer, rather than on SOI samples, making them more promising for integration with the most efficient IBC geometry solar cells. 

Holes of 350-400~nm were opened in a Cr \blue{or \ce{Al2O3}} mask by single laser pulse irradiation, with writing speeds of 10~cm/s.
Smaller holes with 180-245~nm diameter were fabricated in the Cr mask by the second method, stepper projection photo-lithography and lift-off. 
The plasma etching rate of Si vs hole diameter, $W$, in Cr mask was sub-linear. 
Larger holes are vital for faster KOH etching, which is considerably longer $\sim 10$~h as compared with $\sim 10$~min for similar PhC patterns with RIE. Pulses of only 2-5~nJ (515~nm/230~fs) on the sample  focused by a $NA - 0.9$ dry objective lenses are required for 
laser printing of $\sim 400$~nm diameter nano-holes in the Cr-mask\blue{; approximately double the pulse energy was required for opening holes in the \ce{Al2O3} mask}.


\blue{D}ue to strong under etch, the surface of Si where damage occurred during hole ablation is removed by plasma etch. A subsequent KOH or TMAH~\cite{tmah} etch step for Si surfaces can be added for the lowest SRV and best quality at the very end of PhC fabrication. \blue{Use of dielectric \ce{Al2O3} mask is expected to help further reduction of SRV and simplify surface texturing steps.}

Future improvement of large area fs-laser pattering could employ Bessel-like beams to increase tolerance to surface tracking~\cite{13ome1862} and adopt stitch-free laser writing~\cite{19oe15205}. The main advantage of direct laser writing is due to it's vacuum-free mode of operation in mask definition. \blue{Over the last 20 years, the trend for the average power of ultra-short pulsed lasers has been to double every two years, analogous to Moore's law scaling~\cite{moore} as reported by Amplitude Ltd. lasers~\cite{Mottay}, i.e. the number of photons packed in time (power) scales like the transistors per area/volume on a chip. Flexibility of material processing by burst mode ablation~\cite{Kerse,Neuenschwander} has fueled larger-area $> 1$~m$^2$ industrial applications such as hydrophobic and anti-ice surface treatments or injection mold texturing~\cite{Bonaccurso,Hodgson}. Fast polygon scanners with beam travel speeds up to hundreds of meters per second have been implemented to disperse $\sim$kW radiation power. These trends are promising for solar cell texturing on increasingly larger areas by direct laser writing demonstrated in this study.} 


\small\section{Experimental Section}\label{meth}

The energy balance of the absorbed, reflected, and transmitted parts of light energy $A+R+T = 1$ was calculated from the refractive index $\Tilde{n}$ of solar cell material (Si in this study~\cite{Green2008}) at 
the normal incidence as $R = [(n-1)^2 +\kappa^2]/[(n+1)^2 +\kappa^2]$ and $A = 1-R = 4n/[(n+1)^2 +\kappa^2]$. These expressions are valid for light incidence from air $n_{air} = 1$. The Beer - Lambert law defines the transmitted part $T = e^{-\alpha d}$ in a single pass through thickness $d$; $\alpha d\equiv \ln(10)OD$, where the optical density (absorbance $A$) is calculated from the transmittance as $OD = -\lg T$. From the known spectrum of Silicon permittivity $\varepsilon(\lambda) = \Tilde{n}(\lambda)^2$, the efficiency of light harvesting by different Si solar cell thickness was estimated and compared with with the corresponding Lambertian limit. Three different methods to define a light trapping PhC pattern on Si solar cell surface are described next.

\subsection{Femtosecond laser writing} Direct fs-laser writing by ablation of Cr-on-Si mask was carried out with a solid state Yb:KGW laser (Pharos, Light Conversion, Ltd. Vilnius, Lithuania) at the $\lambda = 515$~nm wavelength with $t_p = 230$~fs pulses at repetition rate $f=200$~kHz. High precision mechanical stages with write speed up to 25~cm/s (Aerotech, GmbH. Pittsburgh, USA.) were used for the lateral in-plane (xy) translation of the sample during laser write. Axial (z) positioning of the focal spot onto Si surface was made by a high-precision stage mount of the objective lens. A software-hardware integration solution of the entire fs-fabrication unit was made by Altechna Ltd. Ablation was carried out with focusing using objective lens with numerical aperture $NA = 0.9$ (Mitutoyo) at a low pulse energy $E_p = 2-15$~nJ (on the sample) with pulse-to-pulse separation 
for specified period of PhC from $\Lambda = 1~\mu$m to 3.1~$\mu$m. 
The laser beam profile was measured utilising a Ophir Spiricon SP928 beam profiling camera.

Laser fabrication was carried out in air (cleanroom, class 1000). Crystalline c-Si (n-type; Phosphorus-doped) $\langle100\rangle$ and intrinsic samples (University wafer)
The conductivity and its type of Si was determined by the van der Pauw method with an Ecopia HMS-3000. 
There was no difference observed for the tested low conductivity intrinsic n- and p-type Si in terms of formation of the laser ablation patterns and PhC plasma etching. 

All samples were cleaned in acetone and isopropyl alcohol (IPA) prior and post etching to minimise surface contamination. 
A Cr thin film deposition on Si was made by an electron beam evaporation (Kurt J. Lesker AXXIS system)
A reactive ion etching (RIE) was carrier out with RIE-101iPH (Samco)
equipped with a load-lock system. 
Structural and optical characterisation of laser ablated regions and PhC structures were carried out by the SEM operation mode of EBL writer (Raith EBL 150$^\mathrm{TWO}$).
 
\subsection{Stepper photo-lithography} An i-line (365~nm) stepper NIKON NSR-2205i12D at Nano-Processing Facility (NPF)  AIST, Tsukuba, Japan was used for this study. A reticle mask (inset in Figure~\ref{f-tsukuba}(a)) consisting of four square regions with different PhC period, $\Lambda$, between a square array of $2~\mu$m diameter disks (a light block for positive resist) were designed over the uniform exposure field of $10\times 10$~cm$^2$. The pattern was projected on a 6-inch Si wafer using a $5^\times$ demagnification with repeated exposure using the i-line stepper. Each quarter segment was a $\sim 1\times 1$~cm$^2$ area after the exposure (Figure~\ref{f-tsukuba}(b)). This pattern formed the mask for etching the photonic crystals (PhC). 

An adhesion promoter for photoresist Hexamethyldisilazane (HMDS) was applied on Si wafers by dip coating after which 700~nm of positive resist (PFI-38) was spin coated onto the 6-inch Si substrate.
After pre-baking at 90$^\circ$ for one minute, an exposure of 25 regions over 6-inch Si wafer was carried automatically (step \& repeat; inset in (b)). Different exposure time (sub-1~s) was programmed for each step to optimise the exposure dose ($\sim 200$~mJ/cm$^2$ is required to expose the resist). Post-exposure bake at 110$^\circ$ for 60~seconds was implemented. Development of resist after exposure was carried out for 10~seconds in NMD-3 2.38\% tetramethylammonium hydroxide (TMAH). The unexposed regions were retained on Si after development and the wafer was diced into $2\times 2$~cm$^2$ chips after development (Figure~\ref{f-tsukuba}(a)). Then, e-beam evaporation of 50~nm of Cr was carried out as described in the previous section. By sonication for 30~min in acetone at room temperature, lift-off over the entire chip area was achieved. The mask was designed to obtain 400-nm-diameter pillars after development (corresponding to $2-\mu$m disks on the reticle). The designed value was approximately twice the achievable resolution of 200~nm for the stepper used.

\subsection{Electron beam lithography} A Vistec EBPG5000plusES EBL writer was used for definition of the KOH wet-etch masks. Two approaches were used, the first produced a \ce{Si3N4} mask, the second produced a Cr mask. The first approach used a combination of LPCVD of \ce{Si3N4}, E-Beam Lithography (EBL) and Deep Reactive Ion Etching (DRIE) to transfer the patterns into the \ce{Si3N4}. The fabrication started from growing 250~nm \ce{Si3N4} on both sides of an SOI substrate with 10~$\mu$m-thick device layer (inset of Figure~\ref{f-conc}(c)). Both, plasma enhanced and low pressure chemical vapour deposition (PECVD $\&$ LPCVD) were tested to form the mask layer. The LPCVD was chosen since it provided more homogeneous, low stress, and porous-free films. A positive ZEP520 EBL resist was used for patterning. The array of patterned squares was oriented to align with the Si wafer flat. After development, the EBL exposure left a grid pattern of unexposed ZEP520 outlining the exposed \ce{Si3N4} square areas. The width of the outline made of unexposed ZEP520 resist was optimised in the design step considering the under-etch of Si below the mask. The resulting ZEP520 grid was used as a mask for DRIE of the \ce{Si3N4} layer below. After DRIE and stripping the resist, the final mask consisted in a \ce{Si3N4} grid pattern outlining exposed square areas of Si. Particular care was taken to fully open the patterned areas into the \ce{Si3N4} in order to expose the Si underneath while preventing excessive etching of the Si substrate. After KOH etching, the \ce{Si3N4} mask was removed with HF. The final device was cleaned with Piranha solution. Photo-lithography was used to define back-side locations for eventual Si etching using AZ-family resist followed by RIE etch of \ce{Si3N4} and stripping of the residual resist mask. 

In the second approach, a Cr mask was made by EBL, Cr e-beam evaporation and lift-off. This approach was carried out on 4-inch Si $\langle 100\rangle$ n-type Phosphorous-doped wafers with a resistivity of 1-10~$\Omega$~cm ($111.75~\Omega/\square$ sheet resistance measured by 4-probe method (Jandel RM3000)). The outline of the squares was defined by exposing a 300~nm-thick layer of PMMA resist, leaving after resist development an array of regularly spaced PMMA squares. The patterned grid was oriented to align with the wafer flat. The spaces between the PMMA square patterns were filled with Cr by e-beam evaporation. The excess Cr and unexposed resist was removed during lift-off process with acetone. The result was a grid mask of Cr outlining the exposed Si square areas. After KOH etching, the Cr mask was removed using Cr etchant solution. The final device was cleaned with Piranha solution.


\subsection{Etching} For potassium hydroxide (KOH) etch, a solution was prepared using 2~litres of 30\% KOH which was then saturated with isopropyl alcohol (IPA) (2~cm surface layer). Etching was carried out at $40^\circ$C. The etch time was evaluated using an online calculator~\cite{KOH} for the known pattern to be etched. Level of p-doping strongly affected the etch time. The depth of the inverted pyramids was dependent on the period $\Lambda$ of the light trapping texture. For the base-angle of pyramid $\beta=54.7^\circ$, the depth $H=(\Lambda/2)\times\tan\beta$ and the side-length at the central cross section of a pyramid $s=\frac{\Lambda}{2}/\cos\beta$. For the used $\Lambda = 0.8, 1.3, 2.5, 3,1~\mu$m, the corresponding depth (for ridge-less inverted pyramids pattern) is $h = 0.57, 0.92, 1.77, 2.19~\mu$m (the aspect ratios $\frac{H}{\Lambda}=0.71$). After etching the light trapping pyramidal texture, the front-side was protected by a Protek coating~\cite{Protek} before a back-side etch step through the SOI substrate. 
For thicker silicon ($\sim 300$ $\mu$m), the KOH solution was heated between 60-70$^\circ$C to reduce etch time. 
Characterisation of fabricated structures was carried out using scanning electron microscopy (SEM), Fourier transform IR micro-spectroscopy (VERTEX 70 Burker), and UV/Vis/NIR spectrometer (PerkinElmer Lambda 1050). 

Plasma etching of tee-pee patterns was carried out with \ce{SF6} 50~sccm, \ce{CHF3} 10~sccm, \ce{O2} 10~sccm for 5 to 30~minutes (etch time dictated different structures); the RF ICP was 180~W, bias 0~W, He pressure 2.70~kPa at a process pressure of 2.5~Pa. Etching of black-Si was made by process: \ce{SF6} 77~sccm, \ce{O2} 100~sccm for 5~minutes at RF ICP 300~W, bias 5~W, He pressure 2.70~kPa and process pressure of 2~Pa. 


\bibliography{report}

\begin{acknowledgments}
We are grateful for project support by Nano-Processing Facility (NPF), AIST, Tsukuba, Japan where we were granted access to photo-lithography stepper. This work was granted by ARC DP190103284 "Photonic crystals: the key to breaking the silicon-solar cell efficiency barrier" project. S. Juodkazis is grateful to the visiting professor program at the Institute of Advanced Sciences at Yokohama National University (2018-20) and to Nanotechnology Ambassador fellowship at MCN (2012-19). This work was performed in part at the Melbourne Centre for Nanofabrication (MCN) in the Victorian Node of the Australian National Fabrication Facility (ANFF). We thank Workshop-of-Photonics (WOP) Ltd., Lithuania for patent licence and technology transfer project by which the industrial fs-laser fabrication setup was acquired for Nanolab, Swinburne. We are grateful to Dan Kapsaskis and An Lee for the training at the characterisation facility.      
\end{acknowledgments}

\section*{Supporting Information}\label{sup}
\setcounter{figure}{0}
\makeatletter 
\renewcommand{\thefigure}{S\@arabic\c@figure}
\makeatother
\begin{figure}[h!]
\centering\includegraphics[width=0.9\linewidth,angle=0]{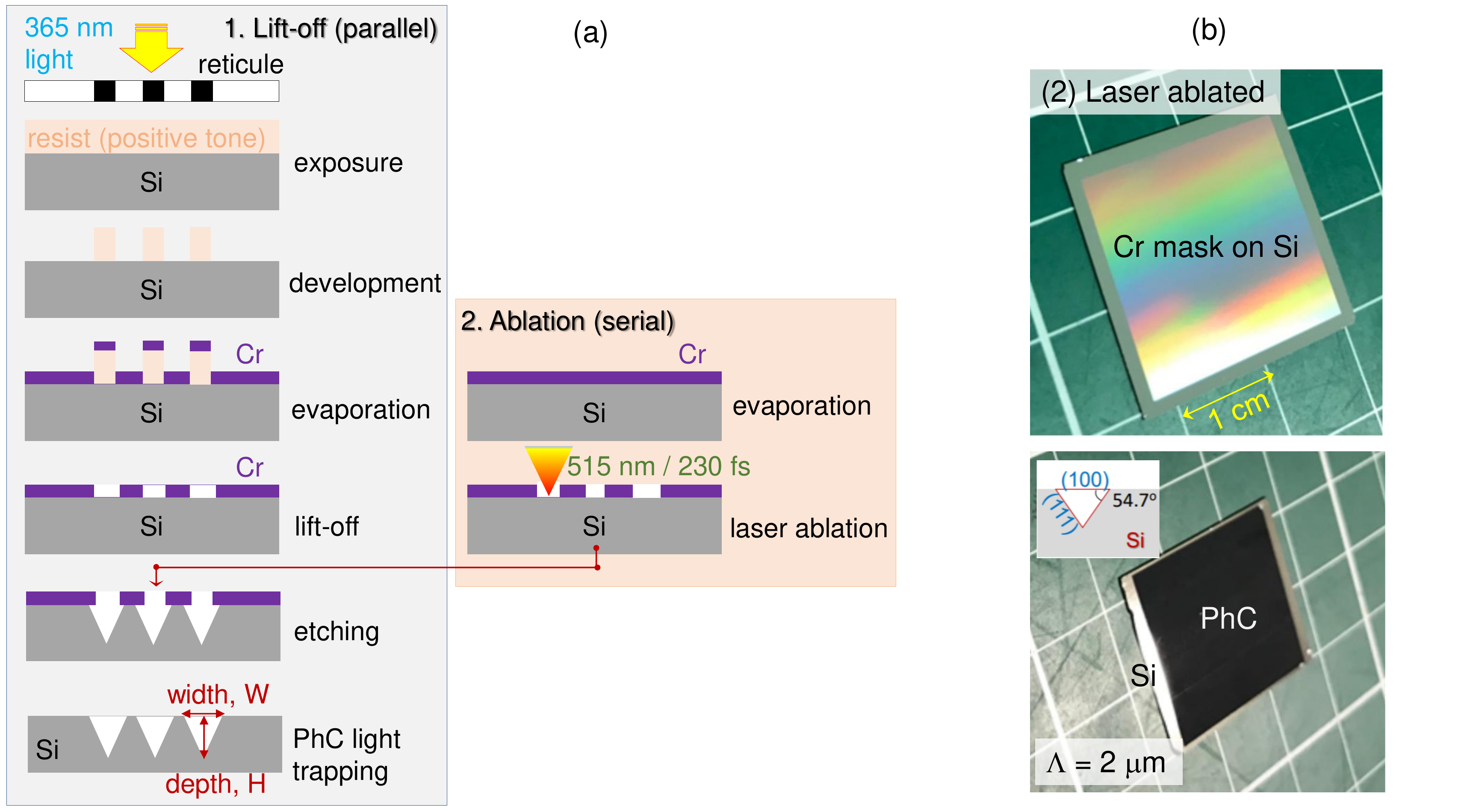}
\caption{(a) A flow chart of processing steps required to fabricate light trapping PhC: 1) lift-off (parallel) and 2) ablation (serial) pathways.  Projection stepper lithography (1) was used with $5^\times$ demagnification. (b) Images of laser ablated mask and the final plasma etched PhC pattern with $\Lambda = 2~\mu$m period. } \label{f-shem}
\end{figure}
\begin{figure}[h!]
\centering\includegraphics[width=1\linewidth]{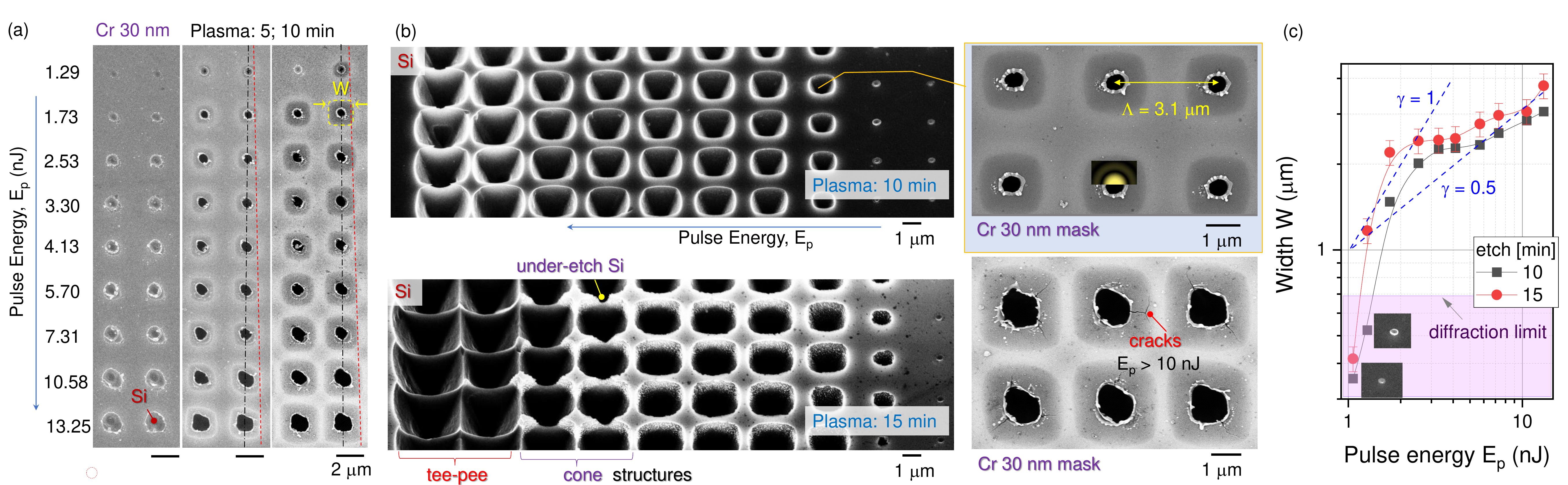}
\caption{Etch through the Cr-mask defined by direct laser writing. (a) SEM images of laser ablated and plasma etched samples.  (b) SEM image of Si after plasma etching for 10 and 15~min and Cr-mask removal; right-side: Cr-mask after plasma etch. (c) Width $W$ evolution for different diameter $D$ of the mask opening ($E_p$ energy); data from (b) where the horizontal measure of the hole was used for the $W$. Plasma etch: \ce{SF6}:\ce{CHF3}:\ce{O2} at 5:1:1 flow rate ratio. The SEM insets shows surface modification for small $E_p$ where Si:Cr alloying occurred. Slope exponents $\gamma = 1, 0.5$ corresponds to the linear and diffusion $\propto\sqrt{D\times time}$ defined processes, where $D$~[cm$^2$/s] is diffusion coefficient, respectively.  } \label{f-plasma}
\end{figure}

\begin{figure}[h!]
\centering\includegraphics[width=1\linewidth]{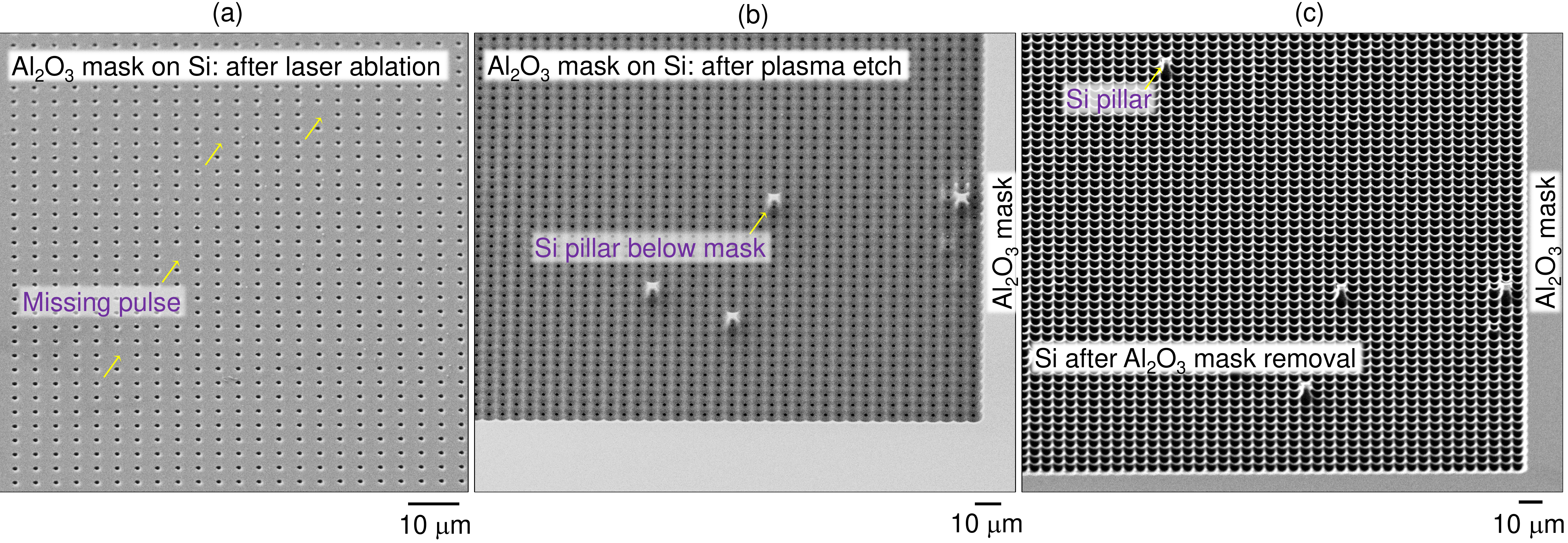}
\caption{\blue{Alumina \ce{Al2O3} (20~nm thickness) on Si as a hard mask for plasma etching; \ce{Al2O3} deposition was made by e-beam evaporation. (a) SEM images of laser ablated holes in \ce{Al2O3} by 515~nm/230~fs pusles of energy $E_p = 12.5$~nJ (at focus). (b) Plasma etched PhC texture at the under-etch conditions (etched protocol was same as for the Cr-mask; Fig.~\ref{f-plasma}). (c) Surface of Si after \ce{Al2O3} mask is removed by ultrasonication in acetone. Upright pyramid is formed with under-etch conditions when an opening in the mask is missing (marked by arrows).}  } \label{f-alumina}
\end{figure}

\begin{figure}[h!]
\centering\includegraphics[width=0.95\linewidth]{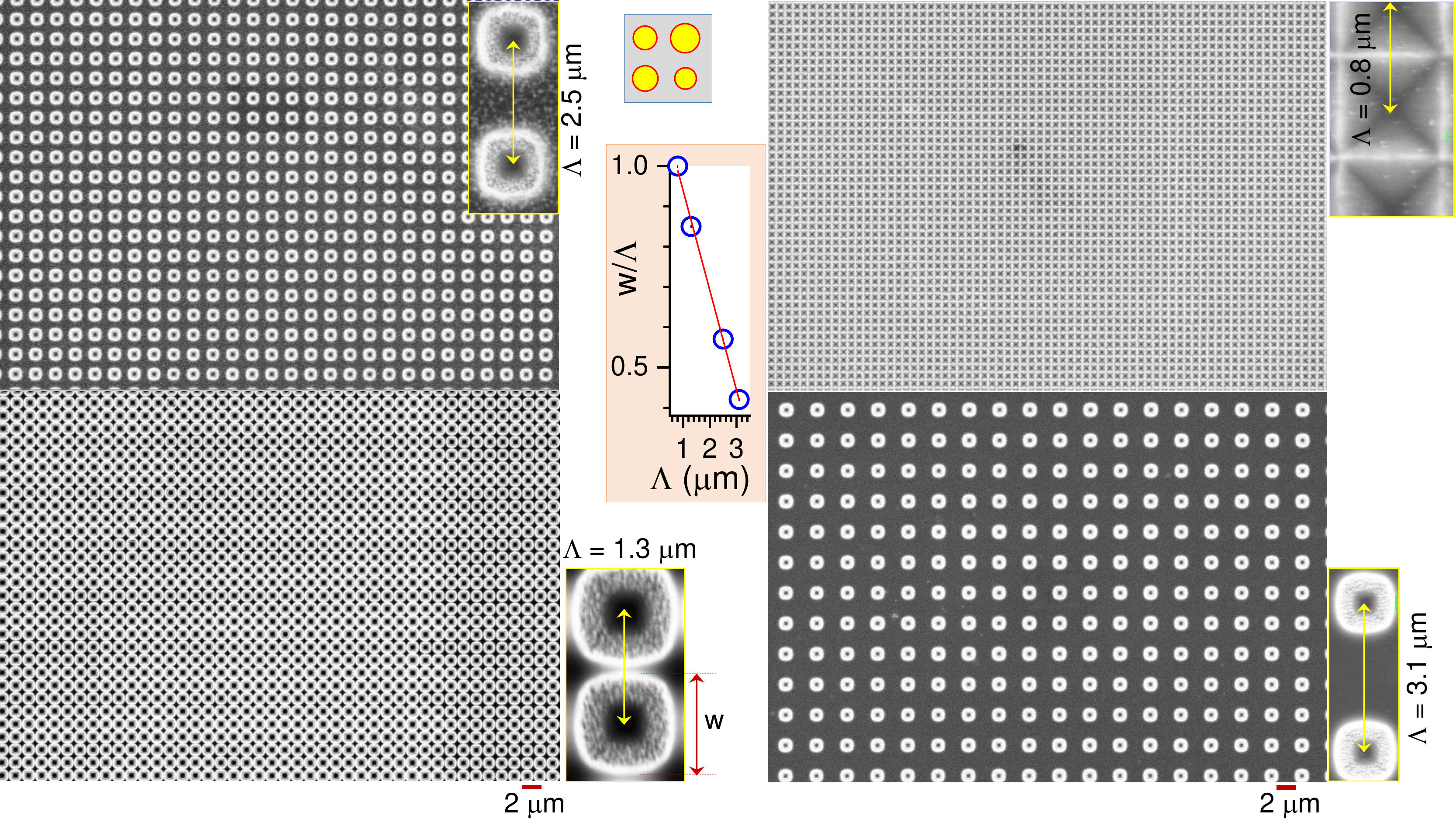}
\caption{Reactive ion etching of PhC patterns \blue{on Si}. SEM images of four-segments after etching and removal of Cr mask (see, Figure~\ref{f-tsukuba} for the geometry). Insets for each pattern shows a closeup view of one period of the \blue{PhC} structure (magnifications of the insets differ). The center plot-inset shows linear scaling of the under-etch ratio $W/\Lambda$ vs. period $\Lambda$. RIE conditions \blue{were}: \ce{SF6} 50~sccm, \ce{CHF3} 10~sccm, \ce{O2} 10~sccm (at bias 5~W) for 7~min.} \label{f-etch}
\end{figure}

\begin{figure}[h!]
\centering\includegraphics[width=1\linewidth]{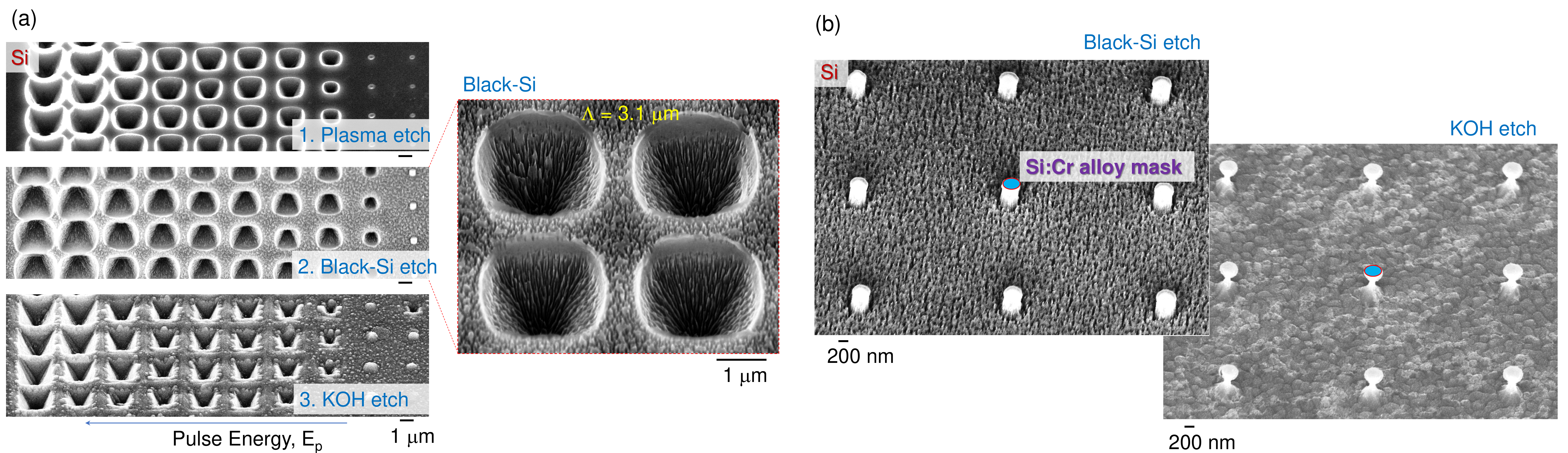}
\caption{Si:Cr alloy formed at low pulse energies $E_p\approx 1$~nJ. (a) SEM images of Si after subsequent etching steps: 1) plasma (as in Figure~\ref{f-plasma}), 2) black Si etch (5~min), 3) KOH  (5-10~min). (b) Formation of Si:Cr alloy at low laser pulse energy irradiation of Cr mask was best revealed by black-Si etch step and subsequent KOH etch. } \label{f-alloy}
\end{figure}

A flow chart of PhC fabrication on Si surface is shown in \textbf{Figure~\ref{f-shem}}(a) and the final result for fs-laser fabrication in (b). Figure~\ref{f-plasma} summarises structural changes during plasma etching through different diameter $D$ holes which were laser ablated through Cr film. Different reactive ion etching (RIE) protocols were tested with a Cr mask with $\sim 200$~nm holes obtained using a stepper lithography followed by lift-off (Figure~\ref{f-tsukuba}).  An optimised RIE procedure to obtain etched pits with the aspect ratio close to one was carried out at flow rates of \ce{SF6} 50~sccm, \ce{CHF3} 10~sccm, \ce{O2} 10~sccm (at bias 5~W) for 7~min; the inductively coupled plasma (ICP) power was 180~W, pressure 5~Pa ($\sim$ 37.5~mTorr). It is noteworthy that the aspect ratio of KOH etch Si is 0.71 for the inverted pyramids on Si(100) surface. \textbf{Figure~\ref{f-etch}} shows SEM images of etched patterns after removal of Cr mask. By defining the opening width of the etched Si pit, $W$, ratio to the period of the pattern, $\Lambda$, it is possible to establish the required etching protocol. Final PhC pattern should have minimum width of the ridges $R$ between adjacent etched inverted pyramids, which corresponds to the condition $W/\Lambda = 1$ or $W+R^{'}=\Lambda$. Inset plot in Figure~\ref{f-etch} shows a linear dependence for the $W/\Lambda = f(\Lambda)$ for the patterns etched in the same run at the etch time of $t_{etch} = 7$~min. The shape of plasma etched inverted pyramids was apparently different (Figure~\ref{f-etch}) resembling KOH-etched shape defined by the crystalline structure of Si for $\Lambda = 0.8~\mu$m while showing tee-pee structure for larger $\Lambda$. 

Figure~\ref{f-alloy} shows results of Si initially etched by RIE then with different dry and wet etching treatments applied. It was observed that with the smallest pulse energy of $E_p\approx 1$~nJ, there were no ablation opening but a noticeable surface modification took place. Plasma etching, which produced a nanotextured black-Si surface was applied to reveal structural modifications on the Cr mask at the low energy laser pulses (see Figure~\ref{f-alloy} (b)). Those regions acted as a mask and prevented surface etching. The mask is expected to be Si:Cr nano-alloy which can have different compositions depending on the formation at the higher temperatures close to the Cr melting at 1860$^\circ$C or that of Si at 1414$^\circ$C~\cite{sicr}. As shown in Section~\ref{res}, the largest part of laser pulse energy $\sim 91\%$ is deposited into Cr. Hence high temperature phases of Cr-rich alloy are expected Cr$_3$Si, $\beta$-Cr$_5$Si$_3$ and $\alpha$-Cr$_5$Si$_3$~\cite{sicr} since they are formed at higher temperatures. However, due to an ultra-fast thermal quenching of the alloy due to a very small volume, formation of non-equilibrium phases of the melt are possible, as was observed for bulk damage of materials with ultra-short laser pulses~\cite{11nc445}. In such a case, Si-rich phases are also possible since they are formed at lower temperatures (closer to the Si-Cr interface): CrSi and CrSi$_2$; interestingly, the latter is a thermoelectric material~\cite{sicr}. Phases such as amorphous-Al$_2$O$_3$, which are not existent at room conditions, can be formed under tightly focused fs-laser pulses~\cite{06am1361}, hence an amorphous Si:Cr nano-alloys can be expected. A separate study is planned to investigate nano-alloy formation at the conditions used.

\newpage
\end{document}